\documentclass[journal]{IEEEtran}

\usepackage{cite}
\usepackage{amsmath,amssymb,amsfonts}
\usepackage{algorithmic}
\usepackage{graphicx}
\usepackage{textcomp}
\usepackage{makecell}
\usepackage{amsmath}
\usepackage{multicol, blindtext}
\usepackage{multirow}
\usepackage{tabularx}
\usepackage{csquotes}
\usepackage{xcolor}
\usepackage{float}
\usepackage{bm}
\usepackage{etoolbox}
\usepackage{gensymb}
\usepackage{threeparttable}

\newcolumntype{C}[1]{>{\centering\arraybackslash}p{#1}}

\graphicspath{ {./figure/} }

\usepackage{nomencl}
\makenomenclature

\usepackage{etoolbox}
\renewcommand\nomgroup[1]{%
  \item[\bfseries
  \ifstrequal{#1}{A}{}{%
  \ifstrequal{#1}{B}{Number Sets}{%
  \ifstrequal{#1}{C}{Other Symbols}{}}}%
]}
\renewcommand{\nompreamble}{\begin{multicols}{2}}
\renewcommand{\nompostamble}{\end{multicols}}

\begin{document}

\title {Monitoring and Control System Development and Experimental Validation for a Novel Extrapulmonary Respiratory Support Setup\\
\thanks{The authors are with the Departments of Mechanical Engineering and Biomedical Engineering at the University of Maryland, College Park, and the University of Maryland School of Medicine in Baltimore, MD, USA. Emails: 
    {\tt\small (mdoostho, karoom, maroom)@umd.edu, (MCulligan, WNaselsky)@som.umaryland.edu, (cthamire, haslach)@umd.edu, (stephen.a.roller, richardhughen)@gmail.com, \\ JFriedberg@som.umaryland.edu, (jhahn12, hfathy)@umd.edu (Phone:+1 301-405-6617)}}%
  \thanks{$^*$Corresponding author.}
}

\author{Mahsa Doosthosseini, Kevin R. Aroom, Majid Aroom, Melissa Culligan, Warren Naselsky,\\ Chandrasekhar Thamire, Henry W. Haslach, Jr., Stephen A. Roller,\\ James Richard Hughen, Joseph S. Friedberg, Jin-Oh Hahn, Hosam K. Fathy$^*$}

\maketitle

\begin{abstract}

This paper presents a novel mechatronic setup intended for providing respiratory support to patients suffering from pulmonary failure. The setup relies upon the circulation of an oxygenated perfluorocarbon (PFC) through the abdominal cavity. Such circulation provides a potential pathway for the transport of oxygen to the bloodstream. However, the viability of this technology for $CO_2$ clearance has not been established. Moreover, there is a lack of experimental data enabling the modeling and identification of the underlying dynamics of this technology. To address these gaps, we develop a flexible experimental perfusion setup capable of monitoring and controlling key variables such as perfusate flowrate, temperature, pressure, and oxygenation. The paper (i) briefly summarizes the design of this setup; (ii) highlights the degree to which its data acquisition system enables the collection and cross-correlation of both perfusion-related and physiological variables; and (iii) discusses the development of flow, pressure, and temperature control algorithms for the setup. Experiments with large animals (swine) show that the setup is capable of successfully controlling the perfusion process, as well as gathering extensive data to support subsequent modeling and identification studies.

 \par

\end{abstract}


\section{Introduction}
This paper addresses the need for the integrated monitoring and control of a novel setup that can augment respiration without interfacing with the blood stream or utilizing the lungs. The setup is intended to address the societal need for providing pulmonary-independent life support to patients with respiratory failure. Respiratory failure is a life-threatening condition whose causes include acute respiratory distress syndrome (ARDS), pulmonary embolism, pneumonia, toxic inhalation, COVID-19 infection, etc. The fact that the U.S. alone has historically seen more than 100,000 ARDS-related hospitalizations annually, even during pre-pandemic years, highlights the public health magnitude of acute pulmonary failure \cite{eworuke2018national}. 

The main functions of the respiratory system are to bring oxygen into the body and expel $CO_2$ out of the body. If either of these two functions, oxygenation or $CO_2$ removal, falls below critical levels, then the patient will not survive without additional support.  If the condition is severe, then the patient will require mechanical ventilation. This is a technique where the airway is intubated with a tube and a balloon that can achieve an airtight seal, allowing positive pressure assistance of the lungs.  Normally, spontaneous respiration is a negative pressure phenomenon. A potential complication of mechanical ventilation, therefore, is barotrauma to the lungs, which can result in ventilator induced lung injury (VILI). VILI can compound the underlying lung dysfunction and exacerbate the pulmonary failure – potentially to a fatal degree~\cite{anzueto2004incidence,aissaoui2014pneumomediastinum,boussarsar2002relationship,gammon1992pulmonary,gammon1995clinical}. 
In such situations, unless gas exchange is augmented by extra-pulmonary means, the patient will not survive.

Extra-corporeal membrane oxygenation (ECMO) is currently the only pulmonary-independent modality available to supplement gas exchange. It involves drawing blood out of the patient through a vascular cannula, oxygenating it, then pumping it back into the patient through another cannula \cite{marasco2008review}. Unfortunately, ECMO is an expensive resource, with one study indicating a mean total hospital cost above \$200,000 per patient\cite{mishra2010cost}. The availability of ECMO is limited by cost and personnel requirements: its initiation is typically performed by specially trained cardiac surgeons, and its maintenance requires constant monitoring by highly trained personnel.
Even when available, ECMO is accompanied by contraindications or exclusion criteria that may make it a nonviable option for patients with potentially reversible lung failure. 
\cite{murphy2015extracorporeal}.
Therefore, there is a need for additional ways to support respiration that do not require the lungs or ECMO.\par

\begin{figure}[h!]
\centerline{\includegraphics[clip, trim=.8cm 3cm .5cm 3cm,width=.5\textwidth]{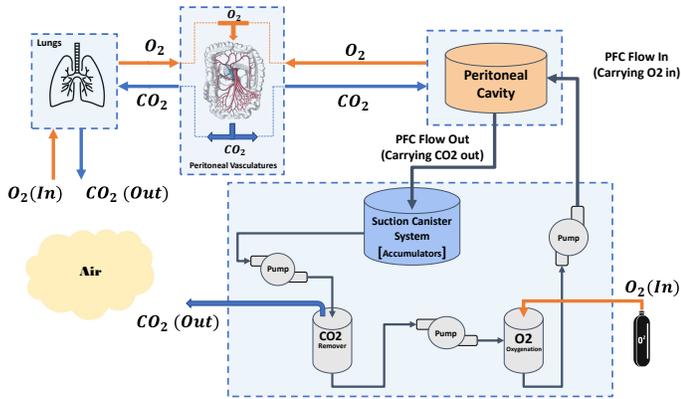}}
\caption{Schematic of the third lung concept}
\label{3rd_lung}
\end{figure}

The main idea motivating this paper is to supplement gas exchange by perfusing (i.e., circulating) an oxygenated perfluorocarbon (PFC) through a patient's peritoneal (i.e., abdominal) cavity. Perfluorocarbons (PFCs) are organic compounds
consisting either predominantly or entirely of carbon and fluorine. They are inert and recognized for their very high oxygen and carbon dioxide solubilities~\cite{riess2005understanding,castro2010perfluorocarbon}. Thanks to these properties, PFCs are well-suited for medical applications \cite{torres2019mini,jahr2020blood,spiess2019perfluorocarbon,bialas2019artificial}. For example, they have been investigated as blood substitutes \cite{cohn2009oxygen,biro1987perfluorocarbon} and are also used for ophthalmologic surgeries \cite{kramer1995perfluorocarbon}. One particularly relevant application is the use of PFCs for \textit{liquid ventilation}, or ``liquid breathing". This refers to filling the lungs partially or completely with an oxygenated PFC in an effort to augment gas exchange\cite{curtis1991cardiac,hirschl1996initial,leach1996partial,hirschl1998partial}. Both laboratory studies and clinical trials have been performed on liquid ventilation. These studies show that while liquid ventilation does indeed supplement gas exchange \cite{shaffer1992liquid,wolfson2004liquid,staffey2008liquid,sarkar2014liquid}, its benefits do not justify its adoption as an alternative to mechanical ventilation~\cite{hirschl2002prospective}. \par

The research in this paper is similar to liquid ventilation in its use of PFC to augment gas exchange, but fundamentally distinct in its use of the abdominal cavity, as opposed to the lungs, for gas exchange. Figure (\ref{3rd_lung}) summarizes this respiratory support approach. A perfusion circuit is used for oxygenating PFC, removing $CO_2$ from it, and warming it to body temperature. The oxygenated PFC is then perfused through the abdomen, where processes such as diffusion allow it to exchange oxygen and $CO_2$ with the bloodstream. Finally, the PFC is drained out of the abdomen, potentially using negative pressure from a suction/vacuum pump. The end result is an innovative approach that allows the peritoneal cavity to be used “like a lung,” analogous to the way it is used “like a kidney” for peritoneal dialysis ~\cite{cullis2014peritoneal}. One potential benefit of this ``third lung" concept is the fact that it offers a pulmonary-independent means of gas exchange that can supplement mechanical ventilation, thereby resting the lungs and helping them heal. Another potential benefit is the fact that the third lung innovation does not require a direct blood-device interface, thereby avoiding many of the risks and contra-indications of ECMO. 

Previous medical research by one of this paper's co-authors (Friedberg) shows that the third lung concept is indeed effective at providing oxygenation to large hypoxic animal models (namely, laboratory swine)~\cite{carr2006peritoneal}. While this prior research is encouraging, it leaves at least four important open questions and research challenges. First, it is not clear what operating conditions (e.g., perfusion flowrates, pressures, temperatures, PFC oxygenation levels, etc.) are ideal for the third lung concept. Second, the impact of the third lung intervention on hemodynamic variables such as heart rate has yet to be fully characterized. Third, the impact of the intervention on $CO_2$ clearance has not yet been examined in the literature. Fourth, there is a need to implement the third lung concept using a mechatronic setup with extensive data acquisition and control capabilities supporting both ongoing animal experiments and potential future human interventions. To address these challenges, the authors have developed a novel mechatronic setup capable of (i) performing controlled peritoneal PFC perfusion experiments and (ii) gathering extensive datasets characterizing these experiments, from both setup-side and physiological sensors. 

The goal of this paper is to present the above mechatronic setup, focusing on its monitoring and control capabilities. The remainder of the paper is organized as follows. Section II summarizes the setup's mechanical design. Section III describes the setup's monitoring and data acquisition system. Section IV summarizes the setup's key closed-loop control functionalities. Section V presents preliminary data from laboratory animal experiments highlighting some of the setup's successes in controlling perfusion parameters such as perfusate flowrate, temperature, pressure, and oxygenation level. Finally, Section VI summarizes the paper's conclusions. 

\begin{figure*}[th]
\centerline{\includegraphics[clip, trim=.1cm 3cm 0.01cm 3cm,width=.8\textwidth]{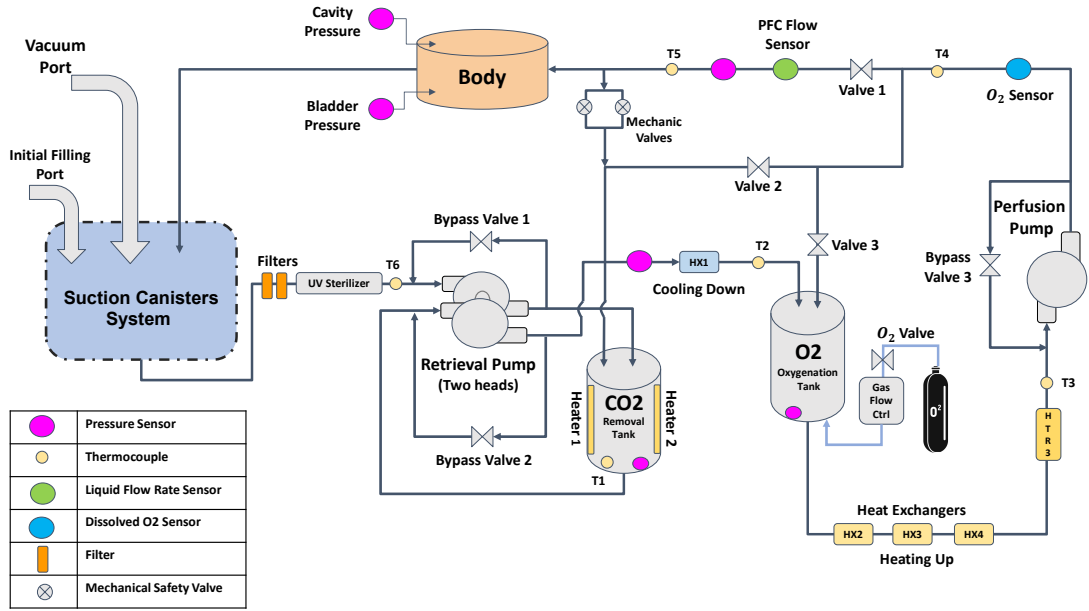}}
\caption{The third lung ventilator setup diagram}
\label{setup_1}
\end{figure*}

\section{Design of the third lung ventilator}
Experiments on the third lung intervention have, to date, focused on large laboratory animals (swine) because they are close in size, core body temperature, and ventilation needs to human patients. The perfusion setup described in this paper is designed to facilitate these experiments, with the ultimate goal of enabling emergency interventions in human patients. The setup is designed to meet five key requirements, namely: 

\begin{enumerate}
   \item Oxygenating the PFC and removing $CO_2$ from it prior to perfusion.
    \item Providing a PFC perfusion flowrate sufficient for supplementing gas exchange. For large animal experiments, prior research suggests that flowrates of up to 6 liters per minute may potentially be required.  
    \item Delivering up to 11 liters of PFC to the abdomen at any given time. This is important, considering the degree to which the abdominal cavity distends during perfusion. Filling the distended abdomen of a 40-50kg adult pig, for example, typically requires 6-7 liters of PFC. 
    \item Achieving perfusion temperatures that are consistent with core body temperature - namely, $37^{\circ}C$ for human patients and $39^{\circ}C$ for laboratory pigs. 
    \item Ensuring safety by avoiding intra-cavity pressures conducive to compartment syndrome. 
\end{enumerate}

\begin{figure}[h!]
\centerline{\includegraphics[clip, trim=.1cm 3cm 0.01cm 3cm,width=.5\textwidth]{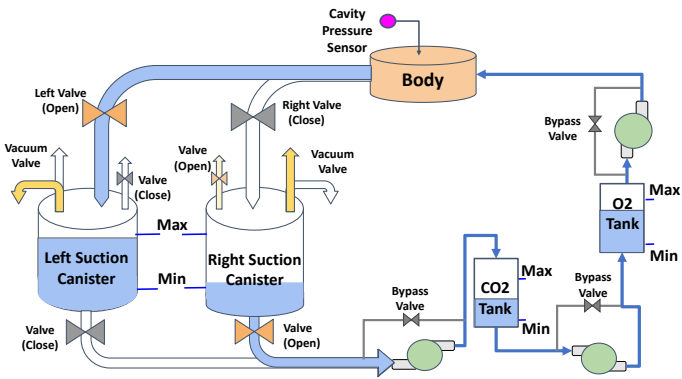}}
\caption{The two-canister accumulator system}
\label{setup_2}
\end{figure}

Figure (\ref{setup_1}) presents the design of the third lung setup, tailored to meet the above requirements. 
When filled, the setup can accommodate 26 liters of PFC, of which 11 liters can be supplied to the animal's or patient's body at any time. The remaining PFC must stay in the setup to ensure that the setup is properly primed and able to sufficiently oxygenate the PFC. PFC enters and leaves the abdomen through tubes typically used as central venuous catheters. Different catheter sizes can be accommodated, a typical size being 36 on the French scale (i.e., $12$mm). An oxygenated PFC (specifically, a mix of cis- and trans-Perfluorodecalin) is perfused through the abdomen of the patient or animal. The PFC then drains into an accumulator using a combination of gravitational drainage and active suction via a vacuum pump. Two different versions of this accumulator have been built and can be rapidly interchanged, namely: a single-canister system and the dual-canister system in Fig. (\ref{setup_2}). The former system uses one canister to receive fluid drained from the test animal and supply it to the rest of the setup. In contrast, the dual-canister system switches periodically between a canister that recovers fluid from the animal versus a canister from which fluid is pumped into the rest of the setup. 

Once the fluid is recovered by the accumulator system, it is filtered then exposed to an ultraviolet flood light. The fluid then passes through a chamber where $CO_2$ is purged. The specific setup sketched in Fig. (\ref{setup_1}) uses PFC heating plus exposure to an oxygen stream as means of $CO_2$ removal, the idea being to rapidly achieve equilibrium between dissolved and incoming gas concentrations. Alternative $CO_2$ removal mechanisms include the use of vacuum to bubble $CO_2$ out of the PFC as well as the use of chemical removal means (e.g., soda lime canisters). If the temperatures used for $CO_2$ removal exceed the ideal perfusion temperature, the setup provides the option to pass the warm PFC through a heat exchanger connected to cold water flow from a cardioplegic heater/chiller. The PFC is then oxygenated using a gas bubble chamber connected to an oxygen tank through an actively-controlled valve, with the possibility that future designs may employ membrane gas exchangers instead. Next, the temperature of the PFC is regulated to meet the desired perfusion target using a mix of electric heating and heat exchange with hot water from the cardioplegic heater/chiller unit. Finally, the oxygenated PFC is pumped into the abdomen. 

As shown in Fig. (\ref{setup_1}), the setup needs to transfer PFC from the suction canisters to the $CO_2$ removal chamber, then to the oxygenation chamber, then finally to the abdomen. Two peristaltic pumps are used for achieving these three functionalities. A dual-head ``retrieval pump" transfers fluid from the suction canister(s) to the $CO_2$ removal chamber, then to the oxygenation chamber. Next, a ``perfusion pump" supplies PFC to the animal. Balancing the PFC fluid levels in the various chambers can be achieved through bypass valves, as shown in Fig. (\ref{setup_1}), with the recognition that modifying the setup to incorporate three independent pumps may potentially provide greater control authority. The setup incorporates a mix of spring-loaded passive and actively-controlled mechanical bypass valves on the final perfusion line. These valves provide the ability to bypass the abdomen if intra-cavity pressure increases beyond critical limits dictated by setup design (in case of the passive valves) or operator input, if the operator dictates a software-based pressure limit (in case of the active valves). This is important for avoiding cavity pressures conducive to compartment syndrome. 

Temperature control is potentially critical for the success of the third lung intervention, and requires significant fluid heating capabilities. Heating is needed for ensuring compatibility between perfusion temperature and core body temperature. Heating beyond core body temperature is also potentially important for $CO_2$ removal. For illustrative purposes, consider the problem of heating the PFC from a room temperature of $22 ^{\circ}C$, to a desired $CO_2$ removal temperature of $42 ^{\circ}C$, assuming a PFC flowrate of $5$ liters per minute. Knowing that the density, $\rho$, of Perfluorodecalin is $1.93~[kg/L]$ and its specific heat capacity, $C_p$, is $1000~[J~kg^{-1}~K^{-1}]$, Eq. (\ref{heat_capacity}) solves for the heat, $Q_{th}$, required for this functionality:  

\begin{equation}
\label{heat_capacity}
\begin{split}
Q_{th} &= \dot{m}c_p\Delta T \\[5pt]
 &= (5~[\frac{L}{min}])*(1.93~[\frac{kg}{L}])*(1000~[\frac{J}{Kg K}])*(20~[K]) \\[5pt]
 &= 193000~[\frac{J}{min}] = \frac{193000}{60~[sec]} \cong 3216~[W],
\end{split}
\end{equation}

\noindent where $\Delta T$ is the desired rise in PFC temperature and $\dot{m}$ is the mass flowrate of PFC. 

The setup is equipped with two $250 [W]$ electric heaters in its $CO_2$ removal chamber plus a $100 [W]$ electric heater attached to the final perfusion line. Moreover, the setup is connected through heat exchangers to the hot and (optionally) cold water outputs of a cardioplegic chiller/heater unit. The chiller/heater unit can provide up to $3000 [W]$ of heat to its output hot water, which can be raised to temperatures as high as $41^{\circ}C$. Given the proximity of this hot water temperature to the final perfusion temperature, three of the setup's four heat exchangers are used for heating the PFC, compared to a single optional heat exchanger for cooling. These details highlight the importance of the coordinated control of the setup's heating (and potential cooling) assets in order to ensure effective perfusion temperature control. Other critical variables that the setup must monitor and control include perfusion flowrate, total perfused volume, perfusate gas concentrations, and intra-cavity pressure. The remainder of this paper presents a detailed description of the monitoring, data acquisition, and control functionalities implemented to meet these goals.  

\begin{figure*}[h!]
\centerline{\includegraphics[clip, trim=0.1cm 3cm 0.02cm 3cm,width=0.77\textwidth]{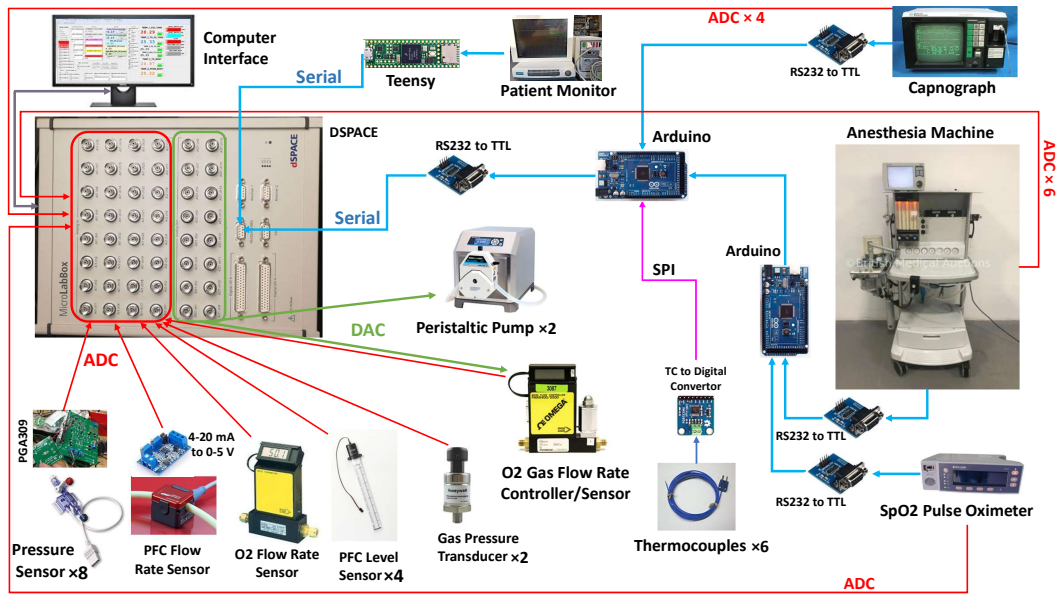}}
\caption{Setup monitoring and data acquisition system}
\label{Signaling}
\end{figure*}

\section{Setup monitoring and data acquisition}

Figure (\ref{Signaling}) provides a high-level overview of the components of the setup's monitoring and data acquisition system. This system: (i) monitors the setup's performance, (ii) monitors the effect of perfusion on the test animal, and (iii) enables important closed-loop control functionalities. Up to 43 signals are collected by this system, from 28 different sensors and patient monitors, in order to collectively satisfy the data acquisition requirements below. Note that the physical locations of these sensors, where possible, are indicated in Fig.s (2-3). 

\begin{itemize}
    \item \textbf{Fluid height/volume monitoring} (6 sensors, 6 signals): The setup has the ability to monitor the height of the PFC in all of its canisters - namely, the oxygenation canister, the $CO_2$ removal canister, and its 1-2 canister accumulator. This is important for controlling the setup, as well as estimating the total volume of fluid delivered to the animal. To achieve this functionality, pressure sensors are mounted at the bottoms of all four canisters. Moreover, two pressure sensors are mounted at the tops of the two suction canisters in the dual-canister accumulator in order to measure air pressure during suction. The difference between air pressure in each canister and the pressure at the bottom of the canister enables the estimation of fluid height in the canister.
    \item \textbf{Secondary fluid height/volume monitoring} (4 sensors, 4 signals): The setup is equipped with an optional redundant method for estimating the height of the fluid in its canisters using 4 fluid level sensors. 
    \item \textbf{PFC flowrate monitoring} (1 sensor, 1 signal): The setup monitors perfusion using a PFC flowrate sensor. 
    \item \textbf{Oxygen flowrate monitoring} (2 sensors, 2 signals): The setup monitors the rate at which oxygen flows into the $CO_2$ removal tank using a gas mass flowrate sensor. This rate is adjusted using a manual valve. The setup also controls oxygen flowrate into its oxygenation chamber using an active gas flowrate controller. This controller provides a measurement of the achieved oxygen flowrate back to the setup's data acquisition system.
    \item \textbf{Oxygen concentration monitoring} (1 sensor, 1 signal): An optical sensor is mounted on the final perfusion line in order to monitor the concentration of dissolved oxygen in the PFC. 
    \item \textbf{Fluid temperature monitoring} (6 sensors, 6 signals): Thermocouples are used for monitoring PFC temperatures at multiple critical points in the setup. Specifically, six thermocouples are used for monitoring PFC temperature at: (i) the $CO_2$ removal tank; (ii) the inlet of the oxygenation tank; (iii) the inlet of the perfusion pump; (iv) the outlet of the perfusion pump; (v) the last point in the setup prior to perfusion; and (vi) either the return flow line or the surface of the final polishing heater, depending on usage of the setup. 
    \item \textbf{Perfusion pressure monitoring} (4 sensors, 4 signals): The setup has the ability to measure the test animal's abdominal intra-cavity pressure and bladder pressure through two catheter-mounted pressure sensors. This is important for ensuring safe perfusion. Moreover, the setup uses two additional pressure sensors for monitoring internal fluid pressures prior to perfusion in order to avoid excessive pumping rates. 
    \item \textbf{Physiological signal monitoring} (4 monitors, 19 signals): The setup has the ability to interface with four different medical monitoring systems in order to collect data regarding the test animal's physiological state (e.g., hemodynamics) and response to perfusion. Specifically, the setup can interface with: (i) a Nellcor \textbf{pulse oximeter}; (ii) a Capnomac \textbf{capnograph}; (iii) a Penlon \textbf{anesthesia machine}; and (iv) a TRAM-RAC \textbf{patient monitor}. Collectively, these devices provide 19 different measurements of physiological and/or anesthesia-related variables, with some of these measurements providing an important degree of redundnacy. For example, pulse oximetry is monitored using both the Nellcor oximeter and the TRAM-RAC patient monitor. 
\end{itemize}

\begin{figure}
\centerline{\includegraphics[clip, trim=3cm 9cm 2cm 5cm,width=0.5\textwidth]{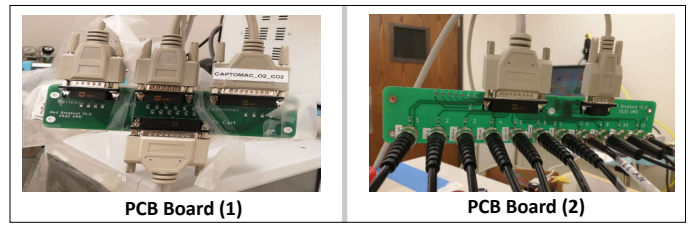}}
\caption{Medical equipment communication boards}
\label{PCBBoards}
\end{figure}

The setup's monitoring and control system is built around a central data acquisition board - in this case, a dSpace MicroLabBox II board. Similar data acquisition boards are often used for instrumentation and control research \cite{ghaffari2012dspace,potnuru2018design}. All analog sensor/monitor signals are read directly by the dSpace board's analog-to-digital converter. For sensor signals that use a current-based (i.e., $4-20mA$) analog communication protocol, standard integrated circuits are used for converting the signals to a voltage-based protocol first, prior to dSpace-based data acquisition. The remaining signals are communicated using either the RS-232 serial protocol or the serial peripheral interface (SPI) protocol. Two Arduino Mega 2560 boards, together with a Teensy board (an Arduino clone) are used for reading these non-analog signals. As shown in Fig. (\ref{Signaling}), one of the Arduino boards is used for aggregating data from the pulse oximeter and anesthesia machine. The second Arduino board adds capnograph and SPI-based temperature data to this aggregate datastream, and forwards it to the dSpace board. Finally, the Teensy board selects specific signals from the fairly large dataset communicated serially by the TRAM-RAC patient monitor, repackages these signals, and communicates them directly to the dSpace board. Minimizing the amount of physical wiring necessary for communicating between these various devices is important, given the space limitation of a typical operating room. To address this issue, the setup collects the signals from all the medical devices using a custom printed circuit board (PCB), as shown on the left hand side of Fig. (\ref{PCBBoards}). A second custom PCB board, shown on the right hand side of Fig. (\ref{PCBBoards}) is then used to input these signals into the Arduino and dSpace boards. 

Different components of the above data acquisition system monitor different underlying dynamics, with significantly different associated time constants. For example, if any of the setup's tubes are accidentally pinched during an experiment, the dynamics of the associated rise in fluid pressure are likely to be much faster than the dynamics of animal blood gas concentrations. With this in mind, the setup's dSpace board has a relatively fast master sampling time of $10ms$, corresponding to a sampling rate of 100 samples/second. Other components of the setup have progressively slower sampling rates and/or data processing times. For example: (i) The SPI protocol is used for reading temperatures every $100ms$. (ii) The Arduino boards communicate data to the dSpace board every $150ms$. (iii) The Teensy board communicates data to the dSpace board every $500ms$. (iv) The pulse oximeter, anesthesia machine, and patient monitor communicate data to the Arduino and Teensy boards every $2000ms$. (v) The capgnograph communicates data to the corresponding Arduino board every $10,000ms$. (vi) Finally, the oxygen sensor updates its readings every $60$ seconds, with the caveat that this is the only sensor that is read by a (proprietary) standalone program not communicating with the dSpace board. 

\begin{figure}[h!]
\centerline{\includegraphics[clip, trim= 2.6cm 11cm 3.5cm 11cm,width=.45\textwidth]{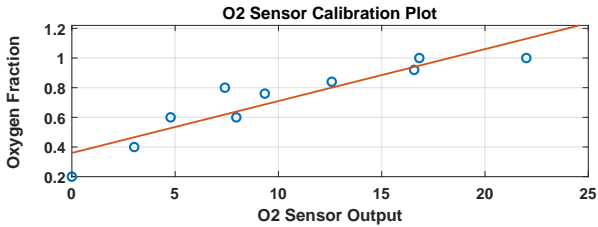}}
\caption{Calibration plot for optical oxygen sensor}
\label{O2_Sensor_Regression}
\end{figure}

Ensuring proper calibration of the above sensors and monitors is essential for the successful use of the setup. Four particular calibration efforts are needed on a regular basis, potentially as frequently as once per use of the setup for animal experiments. First, it is important to calibrate the PFC flowrate sensor, especially if it is used for closed-loop perfusion flowrate control. This is achieved by using the setup to pump a known volume of fluid into an external tank, then calibrating the sensor's data processing software to ensure that the time integral of the flowrate measured by the sensor matches the volume of fluid delivered. Second, it is important to calibrate the canister pressure sensors to ensure correct fluid height estimates. This is done by filling the canisters to known fluid heights, then adjusting the reference voltage outputs of these sensors to furnish height estimates matching these known heights. Third, it is particularly critical to calibrate the cavity and bladder pressure sensors. This is done at the beginning of each animal experiment by exposing these sensors to atmospheric pressures, then adjusting the reference voltage outputs of these sensors to furnish a correct reading of atmospheric pressure. Finally, it is important to calibrate the optical dissolved oxygen concentration sensor. This sensor produces a raw output signal that does not equal dissolved oxygen concentration, but can be correlated to it. To calibrate this sensor, two samples of PFC were prepared with dissolved oxygen concentrations corresponding to partial oxygen pressures of 159mmHg and 760mmHg, respectively, at room temperature. These samples were then mixed in different proportions in order to prepare PFC samples with intermediate oxygen partial pressures. A small but noticeable increase in fluid turbidity occurred at dissolved oxygen partial pressures approaching 760mmHg. The sensor's reading was then correlated to dissolved oxygen fraction, defined as the partial pressure of dissolved oxygen divided by 760mmHg. Figure (\ref{O2_Sensor_Regression}) shows the resulting sensor calibration plot. 

\section{Setup control}

Four closed-loop control functionalities are implemented in the setup. Specifically, the setup contains discrete-event algorithms to control: (i) the filling of the two-canister system when in use; and (ii) the filling of the $CO_2$ removal chamber and oxygenation chambers. The setup also contains proportional-integral (PI) algorithms for controlling (iii) the temperatures of the PFC in the $CO_2$ removal chamber and final perfusion line; and (iv) the final perfusion flowrate/pressure. These controllers are discussed below.
\vspace{24pt}

\begin{center}
    \textit{Multi-Canister Switching Control}
\end{center}

The intent of the dual-canister accumulator is twofold. First, it enables smooth, continuous PFC drainage from the test animal into the setup, potentially in the presence of active suction via a vacuum pump applied to the canisters. Second, it achieves this while minimizing the loading that this suction may apply on the retrieval pump. Discrete-event logic is needed for switching between two configurations. In one configuration, the ``left" canister is receiving drained PFC and the ``right" canister is supplying PFC to the rest of the setup. In the second configuration, these roles are reversed. 

\begin{figure}[h!]
\centerline{\includegraphics[clip, trim= .1cm 10.3cm .1cm 3.5cm,width=.52\textwidth]{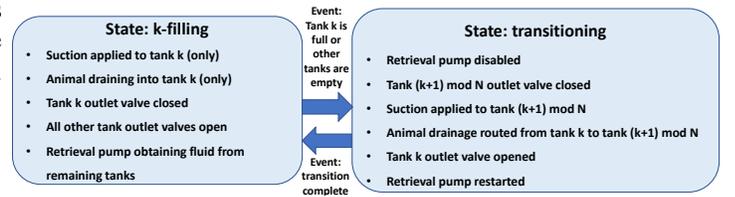}}
\caption{Multi-canister switching control algorithm}
\label{TwoCanisterSwitching}
\end{figure}

Figure (\ref{TwoCanisterSwitching}) summarizes the discrete-event logic used for operating the dual-canister system when it is used. The figure generalizes this algorithm to an $N$-canister system. Most of the time, the setup is in a ``k-filling" state, where canister $k$ is being filled and all other canisters are being emptied. When canister $k$ is full or any other canister is empty, an immediate switch takes place to a ``transitioning" state. This state persists for a fixed duration of time, during which the retrieval pump is shut down, suction is applied to the next canister in the sequence of canisters to fill, fluid is routed to that new canister, the outlet valves from the canisters to the rest of the setup are opened and closed appropriately, and the retrieval pump is restarted. Once this transition is complete, the discrete-event control algorithm automatically moves to a state where it is filling the next canister in the filling sequence, namely, canister number $(k+1) mod N$. 

\begin{center}
    \textit{Control of Oxygenation and $CO_2$ Chamber Filling}
\end{center}

Three different control algorithms/loops are used for controlling the retrieval pump and its associated bypass valves. Collectively, these loops ensure that the oxygenation and $CO_2$ removal chambers are replenished with fluid whenever possible, but prevented from overfilling. 

\begin{itemize}
    \item The first loop controls the retrieval pump flowrate. During normal operation, this flowrate is set to a constant value. However, when both the oxygenation and $CO_2$ removal tanks are full or when all the canisters in the accumulator are empty, the pump enters a temporary shutdown state, where flowrate is set to zero. The pump dwells in this state for a fixed time duration, then returns to normal operation. This translates to a 2-state finite state machine governing the pump's operation, where it transitions automatically from normal operation to shutdown whenever needed. The setup's graphical user interface (GUI) allows the user to define the ``empty" and ``full" fluid levels for all canisters. The GUI also allows the user to dictate the constant flowrate used by the retrieval pump during normal operation. This flowrate should ideally be 0.5lpm-1.0lpm larger than the desired perfusion flowrate to ensure that the oxygenation and $CO_2$ removal tanks are always replenished. 
    \item The second loop controls the bypass valve for the first retrieval pump head. This valve is closed during normal operation, allowing the retrieval pump to supply PFC from the accumulator to the $CO_2$ removal tank. However, when the $CO2$ removal tank is full, this valve automatically switches to a state where it is open, allowing flow to bypass the $CO_2$ removal tank. The valve dwells in this state for a fixed time duration, then automatically returns to the normal operation state. This makes it possible for the retrieval pump to replenish the oxygenation tank through its second pump head, without overfilling the $CO_2$ removal tank. 
    \item The third loop controls the bypass valve for the second retrieval pump head. The logic governing this loop is identical to the logic governing the second control loop, the only difference being the the transition from closed- to open-valve operation is governed by fluid level in the oxygenation chamber. This loop allows the $CO_2$ chamber to be replenished, without overfilling the oxygenation chamber. 
\end{itemize}

\begin{center}
    \textit{PFC Temperature Control}
\end{center}

Two different loops are used for controlling PFC temperature (i) in the $CO_2$ removal chamber and (ii) at the final perfusion point. Both loops rely on proportional integral (PI) control with saturation and anti-windup logic. In both cases, the dynamics of PFC temperature are assumed to be governed by the following simple energy balance: 

\begin{equation}
\begin{split}
    \rho VC_p\frac{dT}{dt} &= \rho QC_p(T_{in}-T) \\ 
    &+hA(T_\infty-T)+uR_oI_o^2
\end{split}
    \label{TempDynamics}
\end{equation}

In the above equation, $T$ is the PFC temperature in the given control volume, assumed to be equal to the control volume's outlet temperature. Depending on the control loop, this control volume is either the $CO_2$ removal chamber or the pipe section/manifold being heated by the final perfusion heater. The volume of PFC being heated is denoted by $V$. Moreover, the density and specific heat capacity of the PFC are denoted by $\rho$ and $C_p$, respectively. Thus, the term $\rho VC_pdT/dt$ equals the rate of change of thermal energy stored in the control volume, assuming that the amount of PFC in this control volume, $V$, is approximately constant. The first term contributing to this rate of change, $\rho QC_p(T_{in}-T)$, equals the rate of energy transfer due to the flow of PFC, where $Q$ is the volumetric flowrate of the PFC and $T_{in}$ is the PFC temperature at the inlet of the control volume. The second term contributing to temperature change, $hA(T_\infty-T)$, represents the rate of convection heat transfer, where $h$ is the heat transfer coefficient, $A$ is the area exposed to convection, and $T_\infty$ is ambient temperature. Finally, the term $uR_oI_o^2$ represents electric heating of the PFC, where $R_o$ is the heater's effective resistance, $I_o$ is the nominal current flowing through the heater when it is turned on, and $u$ is an adjustable pulse width modulation (PWM) duty ratio for the heater. This PWM duty ratio is constrained to have a value between 0 and 1. Note that the above model is used for controlling the setup's two sets of electric heaters, keeping in mind that additional heating/cooling functionalities are provided in an open-loop manner by the setup's heat exchangers.

\begin{figure}[t]
\centerline{\includegraphics[clip, trim=0.01cm 3.5cm .01cm 3cm,width=.5\textwidth]{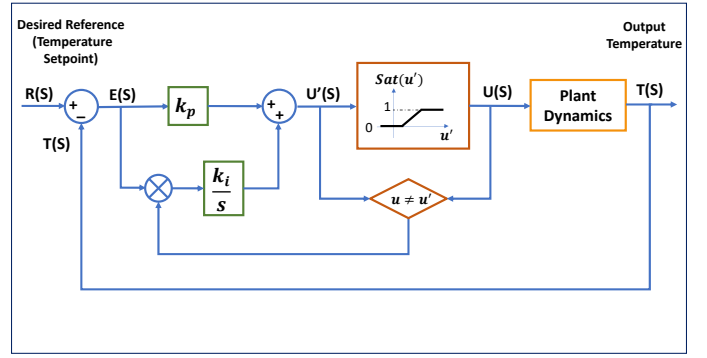}}
\caption{Closed-loop temperature control}
\label{PI_bd}
\end{figure}

Figure (\ref{PI_bd}) shows the PI loop used for temperature control in the $CO_2$ removal chamber. The plant dynamics block in the figure represents Eq. (\ref{TempDynamics}). The difference, $E(s)$ between the desired reference temperature, $R(s)$, and actual temperature, $T(s)$, is passed through a PI controller, $k_p+k_I/s$, with a proportional gain $k_p$ and integral gain $k_I$, in order to produce a PWM ratio $U'(s)$. This PWM ratio is then passed through a saturation function in order to ensure that the final commanded PWM ratio, $U(s)$, lies between 0 and 1. If the saturation function is active, meaning that there is a difference between the signals $U'$ and $U$, then the integral feedback functionality is disabled in order to prevent integrator windup. When the saturation function is inactive, the resulting closed-loop expression for the PFC temperature, $T(s)$, ensures ideal steady-state target temperature tracking in the presence of constant ambient and inflow temperature disturbances, as expected: 

\begin{equation}
        T(s) = \frac{\rho QC_pT_{in}+hAT_\infty+(k_ps+k_I)R_oI_o^2R(s)}{s(\rho C_p(Vs+Q)+hA)+(k_ps+k_I)R_oI_o^2}
    \label{TempControl}
\end{equation}

Temperature control for the $CO_2$ removal chamber was tuned by fitting the open-loop dynamics of Eq. (\ref{TempDynamics}) to an experimental step response test, then using classical pole placement to set the gains $k_p$ and $k_I$. A similar process was used for tuning the gains of the final perfusion temperature controller, with one important caveat compared to the $CO_2$ chamber temperature controller. Specifically, because the final perfusion heater is mounted directly on a metal pipe carrying PFC, as opposed to being immersed in a canister containing PFC, it is significantly more vulnerable to overheating. This vulnerability is particularly noticeable for small or zero PFC flowrates. To address this, the final perfusion heater's controller contains an additional term that brings the corresponding PWM ratio quickly to zero if the temperature of the final perfusion heater exceeds $55^\circ C$. When this function is activated, integral control is disabled in order to prevent integrator windup. 

\begin{center}
    \textit{Perfusion Flowrate and Pressure Control}
\end{center}

The final perfusion controller dictates the flowrate command to the perfusion pump. The controller allows its user to select between two modes: a manual mode and an automatic mode. In both modes, the controller determines a raw flowrate command, $Q'(t)$. In the manual mode, this flowrate command is equal to $k_{ff}Q_{des}(t)$, where $Q_{des}(t)$ is the flowrate dictated by the setup's user through its GUI, and $k_{ff}$ is a feedforward calibration constant determined through setup testing. In the automatic mode, the flowrate command is related to the user-defined desired flowrate as follows: 

\begin{equation}
\begin{split}
    Q'(t) &= k_{ff}Q_{des}(t) \\
    &+ I(t)k_q\int_0^t (Q_{des}(\tau)-Q_{meas}(\tau))d\tau \\
    &+ (1-I(t))k_c\int_0^t (P(\tau)-P_{set})d\tau 
\end{split}
\end{equation}

In the above control law, $P(t)$ denotes peritoneal cavity pressure, and $P_{set}$ is a user-defined pertioneal cavity setpoint pressure that should ideally not be exceeded for prolonged time durations. A dimensionless indicator function, $I(t)$, is defined as being equal to 1 when $P(t)>P_{set}$, and being equal to 0 otherwise. Therefore, when this indicator function equals 1, the implication is that the peritoneal cavity pressure setpoint has been exceeded. The volumetric flowrate, $Q'(t)$, is governed by three terms: a feedforward term identical to the one used for manual flow control, plus two integral feedback terms. The controls gains for these three terms are denoted by $k_ff$ (dimensionless) for the feedforward gain, $k_q$ (units: $s^{-1}$) for the integral flowrate correction gain, and  $k_c$ (units: $m^3Pa^{-1}s^{-2}$) for the integral pressure correction gain. Only one of the above integral feedback terms is active at any given time. When peritoneal cavity pressure exceeds the setpoint pressure, an integral controller with a gain $k_c$ is used for bringing cavity pressure down to the setpoint. In contrast, when peritoneal cavity pressure is below the setpoint, an integral controller with a gain $k_q$ is used for matching the true perfusion flowrate measured by the flowrate sensor, $Q_{meas}(t)$, to the desired flowrate dictated by the user. The actual flowrate command communicated to the perfusion pump is equal to $Q'(t)$ from the above equation, with the exception of two extreme conditions: 

\begin{itemize}
    \item First, when the fluid level in the oxygenation tank drops below a certain minimum level, the perfusion pump controller enters a temporary discrete-event state where perfusate flowrate is curtailed by 50\% while the oxygention tank is replenished. This prevents the excessive emptying of the oxygenation tank. 
    \item Second, when cavity pressure exceeds a user-defined safety limit $P_{max}>P_{set}$, the perfusion pump controller enters a different temporary discrete-event state where perfusion is completely disabled. 
\end{itemize}

The intent of the above perfusion flowrate control algorithms is to ensure steady-state tracking of a user-defined desired perfusion flowrate during normal operation. For safety reasons, this is interrupted when perfusion pressures increase beyond a user-defined setpoint and/or safety limit, or when the oxygenation chamber becomes excessively empty. 

\begin{figure}[tb]
\centerline{\includegraphics[clip, trim=0.1cm 3cm 1cm 3cm,width=.5\textwidth]{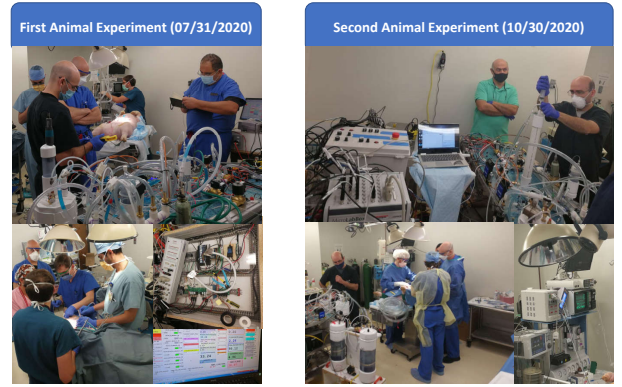}}
\caption{The first and second animal experiment}
\label{test_animal}
\end{figure}

\section{Results}
 
\begin{figure}[tb]
\centerline{\includegraphics[clip, trim= 1cm 13.5cm 1.5cm 7cm,width=.5\textwidth]{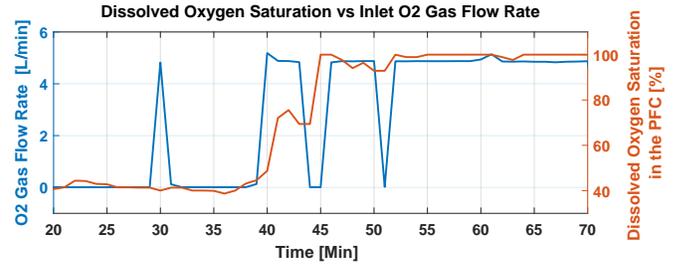}}
\caption{Illustration of setup oxygenation performance}
\label{O2_CTRL}
\end{figure}

 Four animal experiments have been conducted to date, as illustrated in Fig. (\ref{test_animal}). These experiments employed different functionalities in the above setup. For example, the first animal experiment did not employ active suction for drainage of PFC from the test animal, whereas the next 3 animal experiments did. Moreover, the first two animal experiments employed a single-canister accumulator, whereas the third and fourth experiments predominantly employed a dual-canister accumulator. The purpose of this section is to discuss the efficacy of the setup's data acquisition and control functionalities, from a mechatronics perspective. Future work will build on these results, with a deeper focus on the viability of the setup's use for supplementing test animal gas exchange. Seven lessons are visible from experiments performed with the setup to date: 
  
First, the setup is capable of rapidly oxygenating its stored Perfluorodecalin. Figure (\ref{O2_CTRL}) illustrates this by plotting the open-loop commanded flowrate of oxygen into the oxygenation chamber (blue) and the resulting measured percentage of dissolved oxygen in the PFC as a function of time (red), during one of the animal experiments. A rapid increase in dissolved oxygen percentage is achieved while the PFC is recirculated through the setup, in preparation for one of the perfusion events. 

\begin{figure}[b!]
\centerline{\includegraphics[clip, trim=0.01cm 8.5cm 0.001cm 6.6cm,width=.5\textwidth]{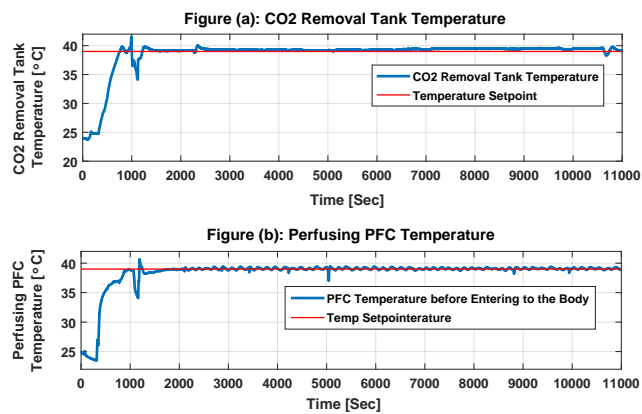}}
\caption{CO2 removal tank and perfusing PFC temperatures (second animal experiment)}
\label{Temp}
\end{figure}

Second, the setup is capable of monitoring and controlling the temperature at which $CO_2$ stripping takes place. Figure (\ref{Temp}) (a) illustrates this, by highlighting the setup's $CO_2$ tank reference temperature tracking performance during a portion of the second animal experiment. Good reference tracking is achieved, with a very small steady-state error corresponding to slight overheating of the PFC. As expected, once the PFC reaches this slightly overheated state, the PWM command to the $CO_2$ tank heaters drops mostly to zero.

Third, the setup is capable of monitoring and controlling the temperature of the PFC at the final perfusion line. Figure (\ref{Temp}) (b) illustrates this by plotting perfusion temperature versus time for a portion of the second animal experiment.

\begin{figure}[h!]
\centerline{\includegraphics[clip, trim= 2cm 13cm 2cm 7.5cm,width=.5\textwidth]{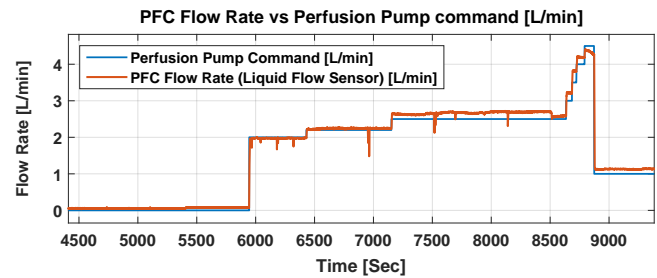}}
\caption{Measured vs. commanded PFC flowrate (third animal experiment)}
\label{PFC_FR_Pump}
\end{figure}

Fourth, the setup is capable of both monitoring and controlling the final perfusion flowrate, in compliance with user input. Figure (\ref{PFC_FR_Pump}) illustrates this by comparing the commanded (blue) versus measured (red) PFC flowrate profiles during a portion of the third animal experiment. The setup was operated in manual flowrate control mode during this particular experiment, as opposed to automated control mode. Therefore, Fig. (\ref{PFC_FR_Pump}) illustrates successful feedforward control tuning, as opposed to successful steady-state flowrate command tracking through integral action. 

\begin{figure}[th]
\centerline{\includegraphics[clip, trim= 1cm 9.8cm 1cm 10cm,width=.49\textwidth]{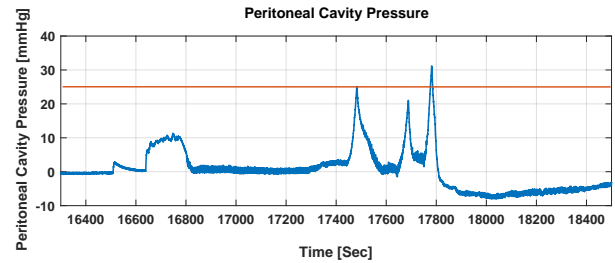}}
\caption{The peritoneal cavity pressure measurement vs the pressure setpoint (Data from the third animal experiment)}
\label{Cavity_P}
\end{figure}

Fifth, the setup is capable of detecting and avoiding excessive peritoneal intracavity pressures. Figure (\ref{Cavity_P}) illustrates this by plotting the setpoint cavity pressure (red) versus measured cavity pressure (blue) for a portion of the third animal experiment. Excursions beyond the setpoint pressure are brief. Moreover, they are followed by rapid curtailment of fluid flowrate (not shown), leading to rapid curtailment of cavity pressure. Negative cavity pressures at the end of the plot are indicative of the use of active suction for fluid retrieval.     

\begin{figure}[t!]
\centerline{\includegraphics[clip, trim=1.8cm 13cm 1.2cm 7.5cm,width=.49\textwidth]{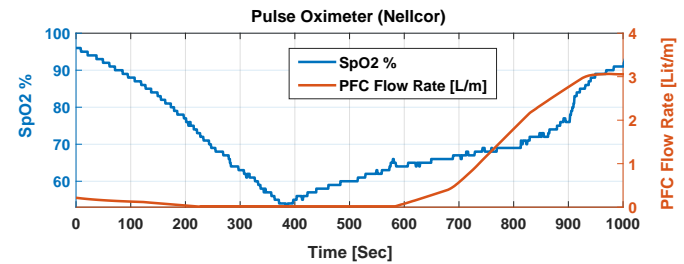}}
\caption{Perfusate flowrate and pulse oximetry (second animal experiment)}
\label{SpO2}
\end{figure}

Sixth, the setup provides sufficient data for assessing the viability of perfusion for oxygenating the test animal. Figure (\ref{SpO2}) illustrates this by showing data from one of the animal oxygenation events. Specifically, the figure plots the animal pulse oximetry (blue) and PFC flowrate (red) versus time. The initial decline in pulse oximetry corresponds to a change in ventilator settings inducing hypoxia. Subsequent improvements in pulse oximetry may be due to a combination of physiological recovery by the animal and/or PFC perfusion. Analyzing the viability of perfusion for improving pulse oximetry is an open topic for ongoing research, exploiting this paper's setup. However, the figure clearly shows that the setup is capable of measuring key variables that can be used for this type of analysis. 

\begin{figure}[b!]
\centerline{\includegraphics[clip, trim=.1cm 9.5cm .1cm 9cm,width=.49\textwidth]{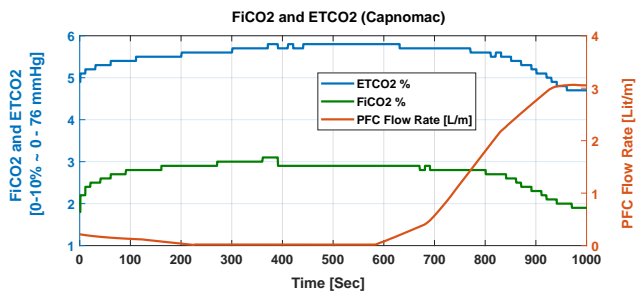}}
\caption{Inspired and end-tidal $CO_2$ concentrations and perfusate flowrate (second animal experiment)}
\label{ETCO2FiCO2}
\end{figure}

Finally, the setup provides sufficient data for assessing the viability of $CO_2$ removal from test animals. Figure (\ref{ETCO2FiCO2}) illustrates this by plotting perfusion flowrate, inspired $CO_2$ concentration, and end-tidal $CO_2$ concentration for a portion of the second animal experiment. Hypercarbia is induced, in this case, through a reduction in minute ventilation. Improvements in end-tidal $CO_2$ concentration (ETCO2) may be caused by physiological recovery mechanisms or perfusion, or a combination of both effects. Analyzing these different recovery mechanisms is left open for ongoing research, building on the setup described in this paper. Please note that the setup's capnograph measures gas concentrations in a tracheal tube, and infers both inspired and end-tidal $CO_2$ concentrations from tracheal measurements. Therefore, changes in inspired gas concentration (FICO2) measurements are likely to reflect gas mixing and re-breathing in the tracheal tube. 

\section{Conclusions and Future Work}

This paper presents the development, design, and implementation of a peritoneal perfusion setup for studies on animal oxygenation using perfluorocarbons. The paper focuses predominantly on the setup's monitoring, data acquisition, and control capabilities as a mechatronics setup. Four animal experiments have been conducted to date using this setup. These animal experiments highlight the setup's functionalities from an engineering perspective. These functionalities include the ability to monitor and control perfusate flowrate, pressure, and temperature. More importantly, the setup is also capable of simultaneously tracking both physical perfusion variables and physiological animal responses. The question of what factors affect the efficacy of peritoneal PFC perfusion as a gas exchange mechanism remain open for ongoing/future research. Moreover, the related question of the minimum viable level of setup complexity and sophistication for gas exchange also remains open for ongoing/future research. The design of the setup in this paper focuses predominantly on achieving a level of sophistication in data acquisition and control that is conducive to scientific exploration, with the recognition that practical clinical implementation may benefit from potential setup simplifications. 

\section*{ACKNOWLEDGEMENTS}

This research was conducted under IACUC \#0121006 at The University of Maryland Medical School, Baltimore, MD (UMB). Support for this research was provided by the Mechanical Engineering Department at The University of Maryland (UMD), UMD startup funding for Dr. Hosam Fathy, an internal grant from The UMD Device Development Fund, and a National Science Foundation (NSF) EAGER grant to Dr. Hosam Fathy, Dr. Joseph Friedberg, and Dr. Jin-Oh Hahn (NSF CMMI Award \#2031251, \#2031245). Any Opinions, findings and conclusions or recommendations expressed in this material are those of the author(s) and do not necessarily reflect those of NSF. The majority of the setup's monitoring, data acquisition, and control functionalities were developed through NSF funding, with funding from other sources contributing to earlier hardware development. The authors gratefully acknowledge all of these sources of support. The authors also gratefully acknowledge the valuable support of other colleagues at UMD/UMB, including Grace Anyetei-Anum, Dr. Hyun-Tae Kim, Dr. Joshua Leibowitz, and Dr. Miao Yu. 
\bibliographystyle{IEEEtran}
\bibliography{Ref}

\vspace{-1.6cm}

\begin{IEEEbiography}
[{\includegraphics[width=.8in,height=1in,clip,keepaspectratio]{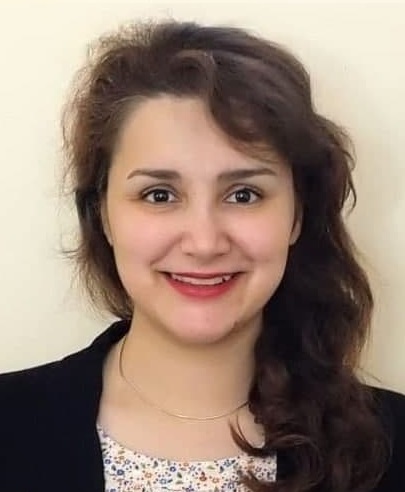}}]
{Mahsa Doosthosseini} received her M.S. (2017) in Mechanical Engineering from the Pennsylvania State University. She is currently a Ph.D. candidate in Mechanical Engineering at the University of Maryland, College Park. Her research focuses on optimal experiment design, control, and mechatronics.
\end{IEEEbiography}

\vspace{-2.1cm}

\begin{IEEEbiography}
[{\includegraphics[width=.8in,height=1in,clip,keepaspectratio]{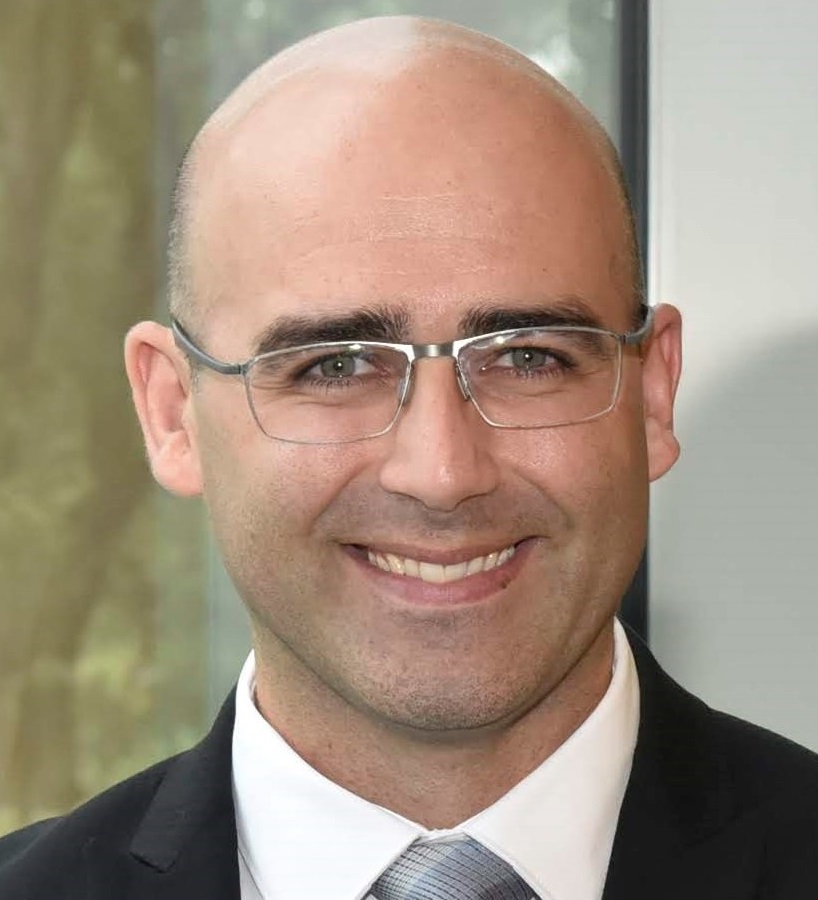}}]
{Kevin R. Aroom} earned his B.S. (2005) and M.S. (2007) in Biomedical Engineering from the University of Maryland, and University of Texas at Austin, respectively. His areas of expertise include rapid prototyping and medical devices.
\end{IEEEbiography}

\vspace{-2.2cm}

\begin{IEEEbiography}
[{\includegraphics[width=.8in,height=1in,clip,keepaspectratio]{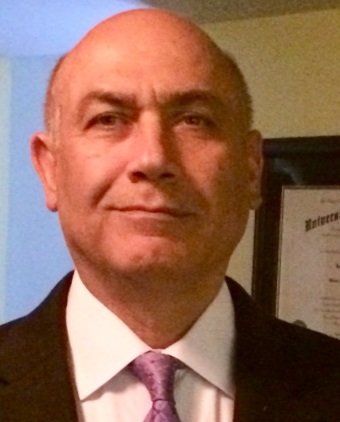}}]
{Majid Reza Aroom} recieved his B.Sc (1984) from the Northeastern University, Boston Massachusetts, USA. He is a project engineer in Thermal Systems, and a Mechanical Engineering Faculty Specialist currently at the University of Maryland.
\end{IEEEbiography}

\vspace{-2cm}

\begin{IEEEbiography}
[{\includegraphics[width=.8in,height=1in,clip,keepaspectratio]{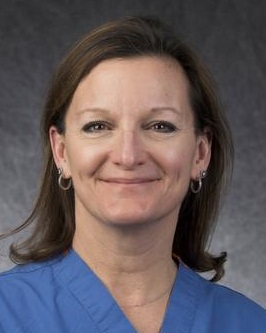}}]
{Melissa Culligan} received her M.S. in Clinical Research Management from Drexel University. She is the Director of Clinical Research at the Division of Thoracic Surgery at The University of Maryland Medical Center, and a Nursing Science Ph.D. student at The University of Maryland School of Nursing. 
\end{IEEEbiography}

\vspace{-2cm}

\begin{IEEEbiography}
[{\includegraphics[width=.8in,height=1in,clip,keepaspectratio]{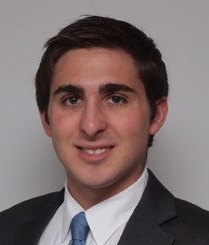}}]
{Warren Naselsky}
earned his B.S. in Biochemistry (2012) and M.D. (2016) at the University of North Carolina at Chapel Hill. He is currently an integrated cardiothoracic surgical resident at the University of Maryland. He recently completed an NIH T32 postdoctoral fellowship at the University of Maryland and now is pursuing a career as an academic thoracic surgeon.
\end{IEEEbiography}

\vspace{-2cm}

\begin{IEEEbiography}
[{\includegraphics[width=.8in,height=1in,clip,keepaspectratio]{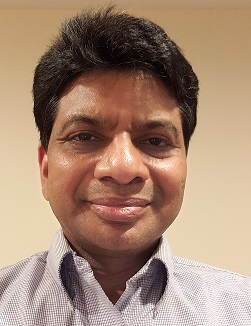}}]
{Chandrasekhar Thamire} holds a B.S. (1987),  M.S. (1997), and Ph.D. (1997) degrees, all in mechanical engineering, and is a licensed engineer. His areas of expertise are turbomachinery design and biological transport. He has worked in industry and academia in various capacities. He is presently a principal lecturer at the University of Maryland.
\end{IEEEbiography}

\vspace{-2cm}

\begin{IEEEbiography}
[{\includegraphics[width=.8in,height=1in,clip,keepaspectratio]{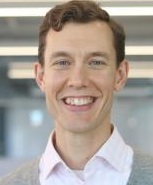}}]
{Stephen A. Roller} holds an MBA from the University of Minnesota Carlson School of Management and has his M.Sc. from the University of Michigan in Biomedical Engineering. He also holds his B.Sc. from the University of Michigan in Mechanical Engineering. He serves as a venture advisor with a focus on medical device technologies currently at the University of Maryland.
\end{IEEEbiography}

\vspace{-1.9cm}

\begin{IEEEbiography}
[{\includegraphics[width=.8in,height=1in,clip,keepaspectratio]{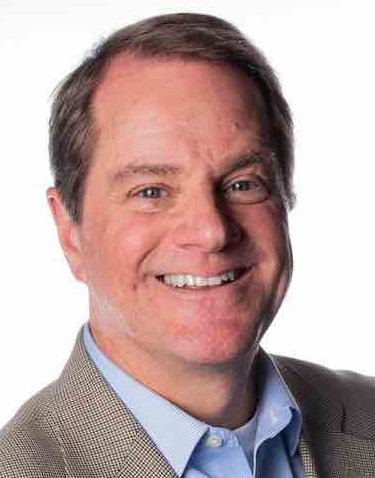}}]
{Richard Hughen} earned his BS and MBA  degrees from The Pennsylvania State University. He is currently CEO of a start-up company developing a predictive continuous respiratory monitoring device for the patient bedside and home. His main expertise is medical device commercialization and is an Entrepreneur in Residence for University of Maryland Ventures.
\end{IEEEbiography}

\vspace{-2cm}

\begin{IEEEbiography}
[{\includegraphics[width=.8in,height=1in,clip,keepaspectratio]{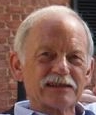}}]
{Henry W. Haslach, Jr} earned a BS in mathematics from Trinity College, an MS in mathematics from the University of Chicago, an MS in Engineering Mechanics and a Ph.D in mathematics both from the University of Wisconsin-Madison. His current research and expertise are on the experimental and analytical rupture mechanics of the human aorta and on blast damage to the brain. He is currently a Research Professor in Mechanical Engineering at the University of Maryland – College Park.
\end{IEEEbiography}

\vspace{-1.7cm}

\begin{IEEEbiography}
[{\includegraphics[width=.8in,height=1in,clip,keepaspectratio]{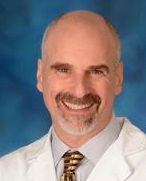}}]
{Joseph S. Friedberg} is the Charles Reid Edwards Professor of Surgery and head of the Division of Thoracic Surgery at the University of Maryland School of Medicine.  He received his bachelor’s degree from the University of Pennsylvania, and received his M.D. from Harvard Medical School.  His interests include pleural mesothelioma, pleural malignancies and disorders, complex lung cancer, chest wall tumors, pulmonary metastasis, malignant pleural effusions, and diaphragm disorders. 
\end{IEEEbiography}

\vspace{-1.6cm}

\begin{IEEEbiography}
[{\includegraphics[width=.8in,height=1in,clip,keepaspectratio]{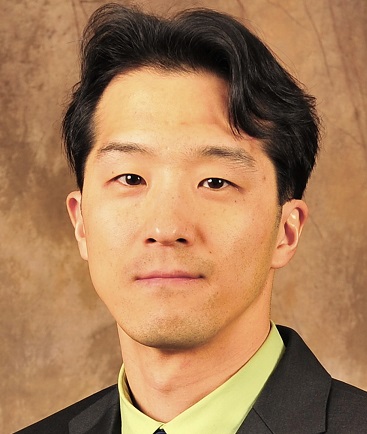}}]
{Jin-Oh Hahn} received BS and MS degrees in mechanical engineering from Seoul National University in 1997 and 1999, and PhD degree in mechanical engineering from Massachusetts Institute of Technology (MIT) in 2008.  He is currently with the University of Maryland, where he is an Associate Professor in the Department of Mechanical Engineering.  His research interests include applications of control, estimation, and machine learning theory to health monitoring, fault diagnostics, maintenance and treatment of dynamical systems with emphasis on health and medicine.
\end{IEEEbiography}

\vspace{-1.3cm}

\begin{IEEEbiography}
[{\includegraphics[width=.8in,height=1in,clip,keepaspectratio]{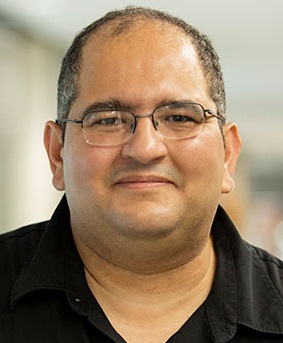}}]
{Hosam K. Fathy} earned his B.Sc. (1997), M.S. (1999), and Ph.D. degrees - all in Mechanical Engineering - from The American University in Cairo, Kansas State University, and The University of Michigan, respectively. His main area of expertise is optimal estimation and control. He is currently a Mechanical Engineering Professor at The University of Maryland.
\end{IEEEbiography}

\end{document}